Revisiting the physical properties of $(LaS)_{1+\delta}(NbS_2)$ misfit-layered compounds

Masanori Nagao[a*], Dan Mouri[a], Yuki Maruyama[a], Akira Miura[b], Yoshihiko Takano[c], and Satoshi Watauchi[a]

[a]*University of Yamanashi, 7-32 Miyamae, Kofu, Yamanashi 400-0021, Japan*

[b]*Hokkaido University, Kita-13 Nishi-8, Kita-ku, Sapporo, Hokkaido 060-8628, Japan*

[c]*International Center for Materials Nanoarchitectonics (MANA), National Institute for Materials Science, 1-2-1 Sengen, Tsukuba, Ibaraki 305-0047, Japan*

[*]Corresponding Author

Masanori Nagao

University of Yamanashi Center for Crystal Science and Technology

Miyamae 7-32, Kofu 400-0021, Japan

E-mail: mnagao@yamanashi.ac.jp

**Abstract**

Electrical transport in polycrystalline and single-crystalline $(LaS)_{1+\delta}(NbS_2)$ misfit-layered compounds was measured. Polycrystalline samples were synthesized using S raw materials of different purities (2N or 6N), and single-crystalline samples were grown using two types of transport agents ($2NH_4Cl+PbCl_2$ or $NH_4Cl$) via the chemical vapor transport method. The temperature dependence on resistivity dropped at ~1.3–2.0 K for some of the samples, which might be affected by the unknown impurity. $(LaS)_{1+\delta}(NbS_2)$ misfit-layered compounds for the main phase of those obtained samples exhibited no superconductivity above 0.2 K by the resistivity measurement.

**Main text**

**1. Introduction**

Misfit-layered compounds exhibit unique physical properties due to incommensurate layered structures [1]. Generally, misfit-layered compounds are composed of monochalcogenide and transition metal dichalcogenide layers, that is, $MX$ and $TX_2$ ($M$ = Sn, Pb, Bi, Sb, rare earth elements; $T$ = transition metal; $X$ = S, Se, Te), respectively [2–4]. The chemical formula of the layer is $(MX)_{1+\delta}(TX_2)$. The incommensurate direction of misfit-layered compounds exhibits no simple periodic structure; therefore, band calculations may be ineffective for predicting physical properties. Some misfit-layered compounds are also superconductors [4–12]. We focused on the $NbS_2$-based ($T$ = Nb, $X$ = S) misfit-layered compounds. This is because $NbS_2$ exhibits superconductivity [13]. For instance, $(SnS)_{1+\delta}(NbS_2)$ and $(BiS)_{1+\delta}(NbS_2)$ misfit-layered compounds become superconductors [7-9,12], but the superconducting transition temperature is lower than that of $NbS_2$. The valence electron states of $M$ (Sn or Bi) elements in these compounds are the s- and p-orbitals (5s + 5p for Sn; 6s + 6p for Bi). On the other hand, the $(LaS)_{1+\delta}(NbS_2)$ misfit-layered compound has a different valence electronic state for the $M$ (La) element, consisting of d-orbital and s-orbital electrons (5d + 6s). Its superconducting transition temperature has been reported to be 2.43 K [11].

However, on the contrary, studies based on resistivity [11] and magnetic susceptibility (and resistivity) measurements [12,14] found that $(LaS)_{1+\delta}(NbS_2)$ was not superconductive above 0.07 K (magnetic susceptibility measurements) [12] and above 4 K (resistivity measurements) [14].

In this study, $(LaS)_{1+\delta}(NbS_2)$ polycrystalline samples were synthesized with varying S raw material purities. Single crystals of misfit-layered compounds are often grown using the chemical vapor transport (CVT) method [14–16]. Thus, single crystals were grown via CVT using each transport agent and raw S material. The determination of superconductivity or not of those obtained $(LaS)1+\delta(NbS2)$ poly-crystalline and single-crystalline samples has measured the electrical resistivity and the magnetic susceptibility (Only single crystals) down to approximately 0.2 K and 2 K, respectively.

## 2. Experimental

The polycrystalline $(LaS)_{1+\delta}(NbS_2)$ misfit-layered compounds were synthesized via a conventional solid-state reaction. The raw materials (total weight 0.5 g), including $La_2S_3$ (99.9% powder, Mitsuwa Chemicals, Japan), Nb (99.9% powder, Kojyundo Chemical Lab., Japan), and S (99.9999%, 6N crystalline, Mitsuwa Chemicals, or 99%, 2N powder, Kojyundo Chemical Lab.), were weighed with a nominal composition of

LaNbS$_3$, which were ground using a mortar, and sealed in an evacuated quartz tube (~10 Pa). The quartz tube was heated at 800 °C for 10 h and spontaneously cooled to less than 30 °C in the furnace. The obtained product was reground and pressed into some pellets of 4 mm in diameter. Similarly, the prepared pellet was placed in an evacuated quartz tube and heat-treated at 1050 °C for 30 h. On the other hand, the single-crystalline samples were grown using the CVT method. The raw materials employed were the same as those for the polycrystalline sample, and the transport agents were NH$_4$Cl (98.5% powder, Kanto Chemical, Japan) and/or PbCl$_2$ (99% powder, Kanto Chemical) for the CVT method. The transport agent was composed of a 2:1 mole ratio of NH$_4$Cl to PbCl$_2$ (i.e., 2NH$_4$Cl+PbCl$_2$) or only NH$_4$Cl. In a previous study [12,14], (NH$_4$)$_2$PbCl$_6$ was used as the transport agent for (LaS)$_{1+\delta}$(NbS$_2$) single-crystal growth. Instead, we used a mixture of 2NH$_4$Cl and PbCl$_2$. The raw materials (total weight of 0.4 g) and transport agents (total weight of 0.02 g) were placed in quartz tubes and sealed in evacuated tubes (~10 Pa). The prepared quartz tubes were heated in a temperature gradient from 1100 °C to 760 °C for 400 h and spontaneously cooled in a two-zone tube furnace. And then, the obtained quart tubes for the poly-crystalline and single-crystalline samples were opened in the air. In previous reports of non-superconducting samples, the sample preparation conditions were as follows. The total weights of the raw materials and $(NH_4)_2PbCl_6$ transport agent were 0.2 g and 0.005–0.01g [12,14], respectively, and the temperature gradient was 930–800 °C [12] or 1110–760 °C [14]. However, the purity of the raw materials was not determined, and details on the preparation of superconducting samples have not been described [11].

The microstructures of the polycrystalline and single-crystalline samples were analyzed using scanning electron microscopy (SEM; Hitachi High-Technologies, TM3030). Their compositional ratios were evaluated using energy-dispersive X-ray spectrometry (EDS; JEOL, EDS equipped JSM-IT700HR) and normalized to Nb (average value) = 1.00. X-ray diffraction (XRD) was to structurally characterize the samples using a Rigaku MultiFlex X-ray diffractometer equipped with a CuK$\alpha$ radiation source. The impurity elements in the S raw materials were qualitatively determined using inductively coupled plasma optical emission spectrometry (ICP-OES); the analysis was performed by Kojundo Chemical Lab., Co., Ltd.

The resistivity–temperature ($\rho$–$T$) characteristics of the obtained samples were measured using the standard four-probe method in constant-current ($J$) mode with a physical property measurement system (PPMS; Quantum Design DynaCool) with an adiabatic demagnetization refrigerator (ADR) option. A 3 T magnetic field was applied to operate the ADR at 1.9 K, and subsequently, the magnetic field was removed.

Consequently, the sample temperature decreased to ~0.2 K. The $\rho$–$T$ measurements commenced at the lowest temperature (~0.2 K), and then the temperature was spontaneously increased to 15 K. The electrical terminals were fabricated with silver paste (DuPont 4922N). Single-crystalline samples were fixed on a single-crystal MgO substrate using silver paste as the electrical terminal. The temperature dependence of the magnetic susceptibility ($M$–$T$) was measured using a superconducting quantum interference device (SQUID) magnetometer (MPMS, Quantum Design) with an applied field ($H$) of 10–100 Oe parallel to the $c$-axis.

## 3. Results and discussion

The polycrystalline samples grown from 2N and 6N S raw materials were black pellets consisting of the $(LaS)_{1+\delta}(NbS_2)$ phase in 00$l$ orientation, as determined using powder XRD (Figure 1). Figure 2 shows the dependence of temperature on resistivity with constant current density ($J$ = 0.05 A/cm$^2$) for the polycrystalline $(LaS)_{1+\delta}(NbS_2)$ samples. The resistivity increased slightly as the temperature decreased. The sample grown from 2N S exhibited a resistivity drop at ~2.2 K, while a small drop was observed at ~0.22 K using the 6N S raw material. None of the samples exhibited superconductivity or zero resistivity.

Four types of single-crystalline samples were prepared with different S raw material purities (2N or 6N) and transport agents (2NH$_4$Cl+PbCl$_2$ or NH$_4$Cl). The obtained crystals were plate-like in shape, with sizes of 1–4 mm and 0.5–1 mm when grown using 2NH$_4$Cl+PbCl$_2$ and NH$_4$Cl transport agents, respectively. More numerous and larger crystals were obtained when PbCl$_2$ was added as the transport agent than when only NH$_4$Cl was used. Crystals were observed at low temperature (760 °C) within the quartz tubes when using 2NH$_4$Cl+PbCl$_2$ as a transport agent. By contrast, when only NH$_4$Cl was used, no crystals were observed at 760 °C; instead, they formed alongside other products at high temperature (1100 °C). NH$_4$Cl decomposed at the heat-treatment temperature, while PbCl$_2$ was maintained in the vapor state during heat treatment, which may have enhanced crystal growth. The crystals grown from the 2N and 6N S raw materials appeared the same. The purity of the S raw materials and transport agents for each single-crystalline sample is shown in Table I. Figure 3 shows the XRD patterns and typical SEM images of a well-developed plane in the crystals grown under each condition, that is, 2NH$_4$Cl+PbCl$_2$ or NH$_4$Cl transport agent and 2N or 6N S raw material purity. The presence of only the 00$l$ diffraction peaks indicates a well-developed $c$-plane corresponding to (LaS)$_{1+\delta}$(NbS$_2$) structures [11]. The $c$-axis lattice constants were 22.92–23.00 Å, which was almost consistent with the reported value of the

(LaS)$_{1+\delta}$(NbS$_2$) phase [11]. Thus, the obtained samples were confirmed to be single crystals based on their XRD patterns. The analytical compositions of the single-crystalline samples are summarized in Table II. Pb was detected in the samples grown using PbCl$_2$ as the transport agent, and the La content was slightly lower than that of the samples without PbCl$_2$. Cl was not clearly detected, and no traces of PbCl$_2$ remained in the grown crystals. A Pb–Nb–S-based compound [17,18] was assumed to exist in small amounts in the obtained samples. Furthermore, compared with the stoichiometric composition of (LaS)$_{1+\delta}$(NbS$_2$), all samples were S deficient, although the La, Nb, and S ratios were nearly consistent with the (LaS)$_{1+\delta}$(NbS$_2$) composition.

Figure 4 shows the temperature dependence of the resistivity along the *c*-plane with constant current density ($J = 1.36$ A/cm$^2$) for the obtained (LaS)$_{1+\delta}$(NbS$_2$) single crystals. The resistivities of all the samples were almost constant above 3 K. The resistivity of the samples using the S (2N) raw materials grown from the transport agents of 2NH$_4$Cl+PbCl$_2$ and NH$_4$Cl dropped at approximately 2 K. It showed at 1.3 K using the S (6N) with 2NH$_4$Cl+PbCl$_2$ transport agent. By contrast, the resistivity decreased slightly for the sample prepared using S (6N) with only NH$_4$Cl. In any case, (LaS)$_{1+\delta}$(NbS$_2$) was revealed to have no superconductivity above 0.2 K, based on the resistivity measurements of the polycrystalline and single-crystalline samples. In particular, the samples synthesized from low-purity S (2N) and/or $PbCl_2$ showed a resistivity drop at ~1.3–2.0 K, which is close to the previously reported superconducting transition temperature determined by resistivity measurements [11]. The compositions of single-crystalline samples after measuring their resistivities were approximately La:Nb:S = 1:1:3, which confirmed the $(LaS)_{1+\delta}(NbS_2)$ misfit-layered compounds by EDS analysis. No correlation between the size/morphology of the obtained single crystals and the resistivity behavior was observed. The resistivity of the obtained samples depended on the purity of the raw S materials; thus, ICP-OES was used to qualitatively analyze the elements in the raw S materials. Only Na and Ca were detected in the range of 30–80 ppm in the S (2N) and S (6N) samples, indicating the presence of the same impurity. This difference was not attributed to elemental impurities in the S raw materials. The raw S materials did not contain other elements that could generally be detected by ICP-OES analysis. Figure 5 shows the temperature dependence of magnetic susceptibility for the obtained $(LaS)_{1+\delta}(NbS_2)$ single crystals. The single-crystalline samples exhibited diamagnetic transitions resembling superconducting transitions. The transition temperatures were approximately 2.3 K and 2.8 K using $NH_4Cl+PbCl_2$ and $NH_4Cl$ as transport agents, respectively. However, the intensities of the diamagnetic transition signals were too low to originate from the main phase. The signals of the $NH_4Cl$ samples were stronger than those of the $NH_4Cl+PbCl_2$ samples. In the quartz tubes, $(LaS)_{1+\delta}(NbS_2)$ single crystals prepared using $NH_4Cl+PbCl_2$ were grown at low temperature (760 °C). For $NH_4Cl$ samples, they were obtained alongside the impurity due to the raw materials in the high temperature region (1100°C), and then the prepared samples for magnetization measurements became susceptible to contamination by impurities from raw materials. Therefore, these diamagnetic transitions originate from impurities. These results indicate that the transition behaviors reflected in the $\rho$–$T$ and $M$–$T$ characteristics were not caused by the Pb–Nb–S-based compound [17,18] observed in the sample synthesized using $PbCl_2$ as the transport agent. We suspect that the impurity causing these transitions in the $\rho$–$T$ and $M$–$T$ characteristics is $Na_x(H_2O)_yNbS_2$ [19] owing to Na in the S raw materials, as detected in the ICP-OES analysis. The other impurity is $Nb_3S_4$ [20], which can be formed from the raw materials of all the samples. Both $Na_x(H_2O)_yNbS_2$ and $Nb_3S_4$ become superconductors at 1.3–3.1 K and 3.78 K, respectively. As a result, we concluded that these additional transitions in the $\rho$–$T$ and $M$–$T$ characteristics did not originate from the $(LaS)_{1+\delta}(NbS_2)$ main phase.

Neither superconducting compounds such as $(SnS)_{1+\delta}(NbS_2)$, $(PbS)_{1+\delta}(NbS_2)$, and $(BiS)_{1+\delta}(NbS_2)$, nor elements such as Sn, Pb, and Bi, were detected. $(SnS)_{1+\delta}(NbS_2)$ [9],

(PbS)$_{1+\delta}$(NbS$_2$) [17], and (BiS)$_{1+\delta}$(NbS$_2$) [7] misfit-layered superconductors of the valence electron state for *M*-site elements (Sn, Pb, and Bi) consist of the s- and p-orbitals. By contrast, non-superconductive (LaS)$_{1+\delta}$(NbS$_2$) is composed of d- and s-orbitals. Therefore, the d-orbital valence electrons may strongly suppress NbS$_2$ superconductivity. However, theoretical studies, such as *ab initio* electronic structure calculations, are required for substantive research.

## 4. Conclusion

Polycrystalline (LaS)$_{1+\delta}$(NbS$_2$) misfit-layered compounds were synthesized using S raw materials of different purities (2N or 6N), and single crystals were grown using either 2NH$_4$Cl+PbCl$_2$ or NH$_4$Cl as the transport agent using the CVT method. The obtained samples were confirmed to be in the (LaS)$_{1+\delta}$(NbS$_2$) phase by the XRD and EDS analyses. Some obtained samples showed a resistivity drop at ~1.3–2.0 K and a diamagnetic transition at ~2.3–2.8 K. Na and Ca were detected as impurities in the samples synthesized from 2N and 6N S. We assume that the transition behaviors in the $\rho$–$T$ and *M*–*T* characteristics originated from Na$_x$(H$_2$O)$_y$NbS$_2$ and/or Nb$_3$S$_4$ superconductors. The main phases in all the prepared samples were (LaS)$_{1+\delta}$(NbS$_2$) misfit-layered compounds, which have no superconductivity. Therefore, we determined that $(LaS)_{1+\delta}(NbS_2)$ misfit-layered compounds were not superconductive above 0.2 K.

**Acknowledgments**

This work was supported by JSPS KAKENHI (Grants-in-Aid for Scientific Research (B) and (C): grant numbers 24K01578 and 23K03358). We would like to thank Editage (www.editage.jp) for English language editing.

Table I Purity of S raw materials and transport agents for each single-crystalline sample.

| Single-crystalline sample | Purity of S raw material | Transport agent |
|---|---|---|
| S(2N)+NH$_4$Cl+PbCl$_2$ | 2N | 2NH$_4$Cl+PbCl$_2$ |
| S(2N)+NH$_4$Cl | 2N | NH$_4$Cl |
| S(6N)+NH$_4$Cl+PbCl$_2$ | 6N | 2NH$_4$Cl+PbCl$_2$ |
| S(6N)+NH$_4$Cl | 6N | NH$_4$Cl |

Table II Analytical compositions of the prepared single-crystalline samples. Normalized to Nb (average value) = 1.00.

| Single-crystalline sample | Analytical composition | | | |
|---|---|---|---|---|
| | La | Nb | S | Pb |
| S(2N)+NH$_4$Cl+PbCl$_2$ | 0.92±0.01 | 1.00±0.01 | 2.90±0.02 | 0.130±0.005 |
| S(2N)+NH$_4$Cl | 0.99±0.02 | 1.00±0.03 | 2.88±0.03 | --- |
| S(6N)+NH$_4$Cl+PbCl$_2$ | 0.90±0.01 | 1.00±0.01 | 2.87±0.02 | 0.132±0.005 |
| S(6N)+NH$_4$Cl | 0.98±0.02 | 1.00±0.02 | 2.83±0.02 | --- |

**References**

[1] G. A. Wiegers, Prog. Solid State Chem. **24** (1996) 1-139.

[2] Y. Gotoh, M. Onoda, K. Uchida, Y. Tanaka, T. Iida, H. Hayakawa, and Y. Oosawa, Chem. Lett. **18** (1989) 1559-1562.

[3] C. Heideman, N. Nyugen, J. Hanni, Q. Lin, S. Duncombe, D. C. Johnson, and P. Zschack, J. Solid State Chem. **181** (2008) 1701-1706.

[4] R. Sankar, G. Peramaiyan, I. P. Muthuselvam, C.-Y. Wen, X. Xu, and F. C. Chou, Chem. Mater. **30** (2018) 1373-1378.

[5] H. Bai, X. Yang, Y. Liu, M. Zhang, M. Wang, Y. Li, J. Ma, Q. Tao, Y. Xie, G. Cao, and Z. Xu, J. Phys.: Condens. Matter **30** (2018) 355701.

[6] A. Nader, A. Briggs, A. Meerschaut, and A. Lafond, Solid State Commun. **102** (1997) 401-403.

[7] A. Nader, A. Briggs, and Y. Gotoh, Solid State Commun. **101** (1997) 149-153.

[8] M. Nagao, A. Miura, Y. Horibe, Y. Maruyama, S. Watauchi, Y. Takano, and I. Tanaka, Solid State Commun. **321** (2020) 114051.

[9] M. H. V. Maaren, Phys. Lett. A **40** (1972) 353-354.

[10] R. Shu, M. Nagao, C. Yamamoto, K. Arimoto, J. Yamanaka, Y. Maruyama, S. Watauchi, and I. Tanaka, J. Alloys Compd. **978** (2024) 173486.

[11] A. Meerschaut, P. Rabu, J. Rouxel, P. Monceau, and A. Smontara, Mat. Res. Bull. **25** (1990) 855-861.

[12] D. Reefman, J. Bank, H. B. Brom, and G. A. Wiegers, Solid State Commun. **75** (1990) 47-51.

[13] C. Witteveen, K. Górnicka, J. Chang, M. Månsson, T. Klimczuk, and F. O. von Rohr, Dalton Trans. **50** (2021) 3216-3223.

[14] G. A. Wiegers and R. J. Haange, J. Phys.: Condens. Matter **2** (1990) 455-463.

[15] W. Y. Zhou, A. Meetsma, J. L. de Boer, and G. A. Wiegers, Mat. Res. Bull. **27** (1992) 563-572.

[16] Y. Gotoh, J. Akimoto, M. Goto, Y. Oosawa, and M. Onoda, J. Solid State Chem. **116** (1995) 61-67.

[17] L. Schmidt, Phys. Lett. A **31** (1970) 551-552.

[18] A. Meerschaut, L. Guemas, C. Auriel, and J. Rouxel, Eur. J. Solid State Inorg. Chem. **27** (1990) 557-570.

[19] A. Lerf, F. Sernetz, W. Biberacher, and R. Schöllhorn, Mater. Res. Bull. **14** (1979) 797–805.

[20] H. Okamoto, H. Taniguti, and Y. Ishihara, Phys. Rev. B **53** (1996) 384-388.

**Figure captions**

Figure 1. XRD patterns of $(LaS)_{1+\delta}(NbS_2)$ polycrystalline samples synthesized using 2N and 6N S raw materials.

Figure 2. Temperature dependence of the resistivity for $(LaS)_{1+\delta}(NbS_2)$ polycrystalline samples using 2N and 6N S raw materials.

Figure 3. XRD patterns and typical SEM images of a well-developed plane for the single crystals grown under each condition listed in Table I).

Figure 4. Temperature dependence of the resistivity ($\rho$–$T$) along the $c$-plane for single crystals grown using each condition given in Table I.

Figure 5. Temperature dependence of magnetic susceptibility ($M$–$T$) for the single crystals grown under zero-field-cooling (ZFC) and field-cooling (FC) conditions (Table I).

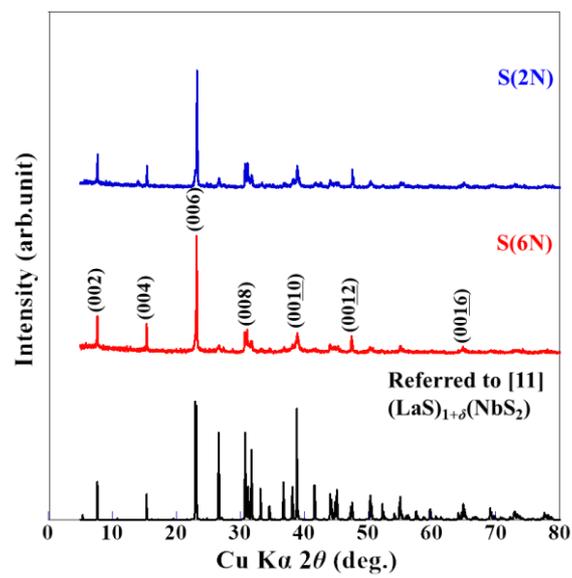

**Figure 1**

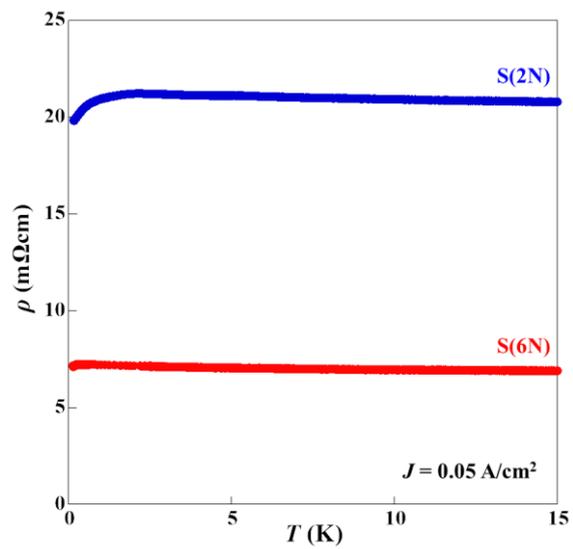

**Figure 2**

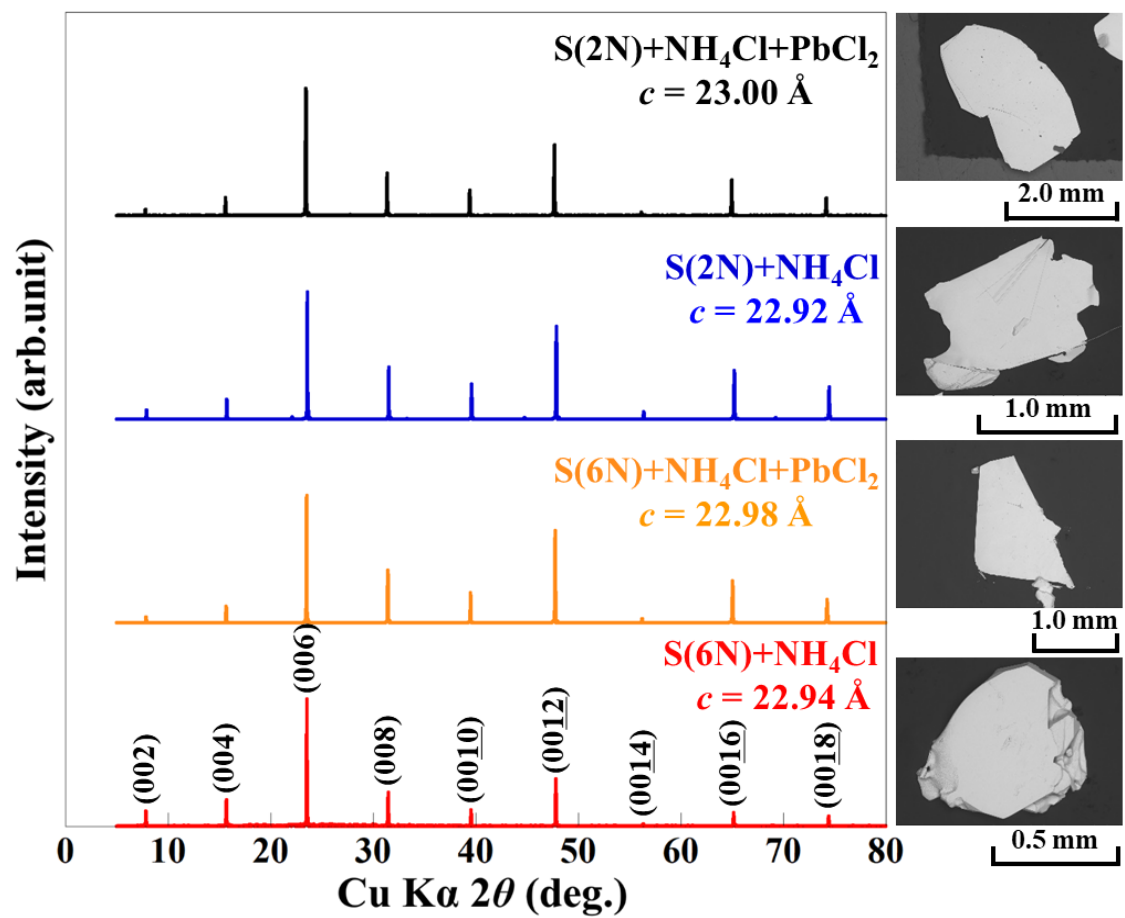

**Figure 3**

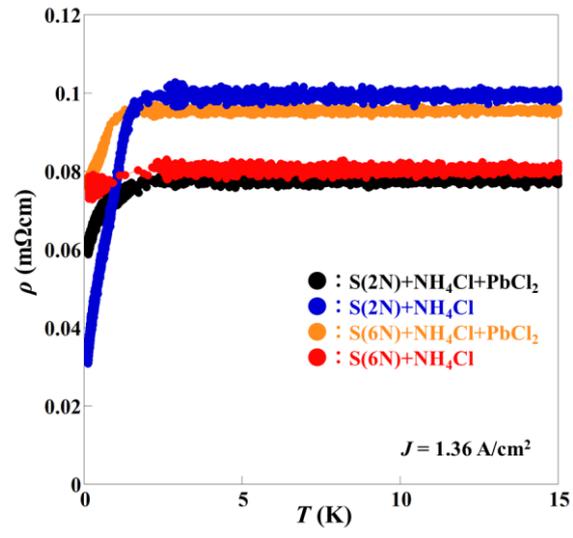

**Figure 4**

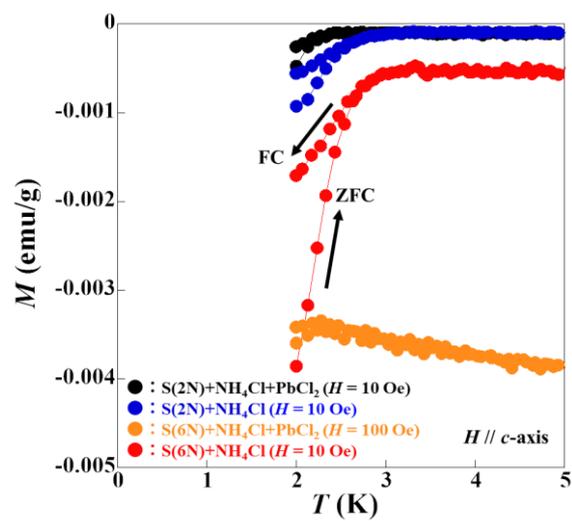

**Figure 5**